\documentclass[a4paper]{jpconf}
\usepackage{graphicx}

\usepackage{cite}
\usepackage{wrapfig}
\usepackage{graphicx}
\usepackage{amssymb}
\usepackage{amsfonts}
\usepackage{amsmath}
\usepackage{longtable}
\usepackage{rotating}
\usepackage{lscape}
\usepackage{epsfig}
\usepackage{multirow}

\usepackage[svgnames]{xcolor}

\usepackage{lineno}

\usepackage[percent]{overpic}

\usepackage{hyperref}
\usepackage[all]{hypcap}
\hypersetup{
    colorlinks,
    citecolor=blue,
    filecolor=blue,
    linkcolor=black,
    urlcolor=blue
}

\def\pt{p_{\rm T}}
\def\av#1{\langle #1 \rangle}
\def 	\r2{\rho_2}

\def\sNN{\mbox{$\sqrt{s_{_{\rm NN}}}$}}

\begin{document}
\title{Reconstruction of  moments of particle distributions with Identity Method at MPD
}

\author{Igor Altsybeev, Vitalii Petrov}

\address{Saint-Petersburg State University, Universitetskaya nab. 7/9, St. Petersburg, 199034, Russia}

\ead{i.altsybeev@spbu.ru}

\begin{abstract}
Precise determination of the moments of multiplicity distributions of identified particles
could be challenging due to the misidentification in detectors. 
The so-called Identity Method allows one to solve this problem. 
In this contribution,  performance of the Identity Method was tested on  
 the A--A events simulated in the conditions of 
the MPD experiment at NICA. With this method, moments within a single kinematic window as well as coefficients of forward-backward pseudorapidity correlations are extracted.
\end{abstract}

\section{Introduction}

Collisions of relativistic nuclei can produce matter at extremely high temperatures and densities. Studies of the transition between the hadronic and partonic phases of this matter are being performed, in particular, with 
various combinations of moments of multiplicity distributions,
for example, by studying event-by-event fluctuations of net-proton number.
This requires identification of different particle species (pions, kaons, protons, electrons).
However, precise determination 
of the moments
can be difficult due to  misidentification in detectors,
for example, due to  overlaps of energy loss ($dE/dx$) distributions
in a Time-Projection Chamber (TPC).
The so-called Identity Method (IM) \cite{IM_2011_first, IM_2011_Gorenstein, IM_2012_Rustamov_Gorenstein}  
allows one to solve this problem
by unfolding the moments of the measured multiplicity distributions for each particle species.
In IM, ``proxies'' for particle multiplicities $W_j$ in each event are constructed as
$W_j=\sum_{i=1}^N  \rho_j(x_i)/\rho(x_i)$, 
$\rho(x_i)=\sum_{j} \rho_j(x_i)$,
where
$j$ denotes a particle type,
$x$ is a value of $dE/dx$
for a given track $i$ (out of $N$ tracks in an event),
and $\rho_j(x_i)$ is the 
$dE/dx$ distribution of particle type $j$
within a given phase space bin.
In contrast to cut-based approach that utilizes  signals in TPC and Time-of-Flight detectors,
the IM allows one to calculate moments keeping high efficiency of particle registration, which is especially crucial for analysis of higher moments.
The IM was  recently used in ALICE for studies of $\pi$, K, p  
\cite{ALICE_rel_fluct_2019} and  net-proton \cite{ALICE_net_proton_2019} fluctuations.

In this contribution,  performance of the Identity Method was tested 
using  data simulated in the conditions of 
the Multi-purpose detector (MPD) at NICA \cite{MPD}, 
which currently is under construction in Dubna.
The MPD will have the TPC, similarly to STAR and ALICE detectors. 
In Section 2, details of the simulation, event and track selection are provided.
In Section 3, the IM is applied for reconstruction of particle momenta in a single (pseudo)rapidity window,
which is the typical use-case of the method.
In Section 4, it is proposed to apply the IM to the studies of the forward-backward rapidity correlations \cite{FB_Capella, Dumitru_2008},
and results of a performance test at the MPD are shown.
The implementation of the Identity Method, used for the current study,
is based on the code available at \cite{IM_2019_Mesut_Anar_code}.

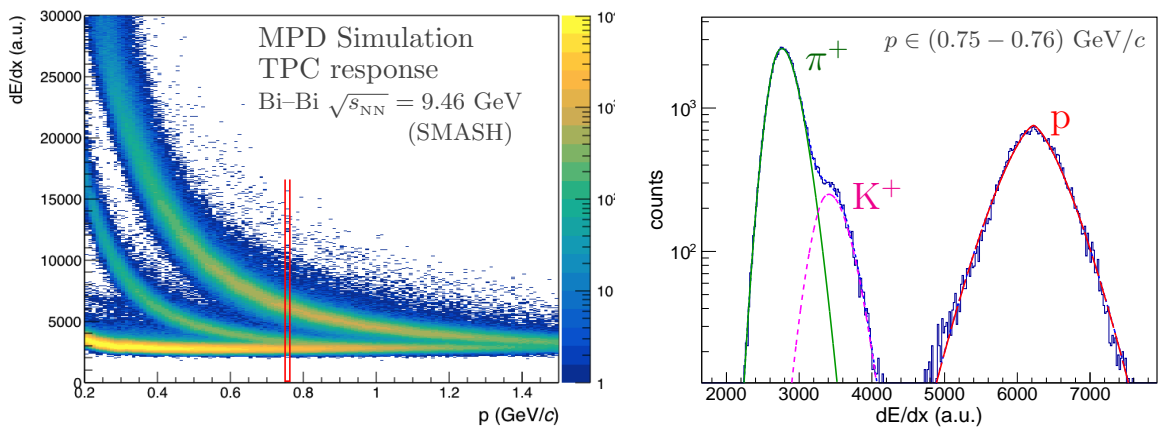
\begin{figure}[t]
\centering
\begin{overpic}[width=0.51\textwidth, trim={0.1cm 0.0cm 0.3cm 0.3cm},clip]{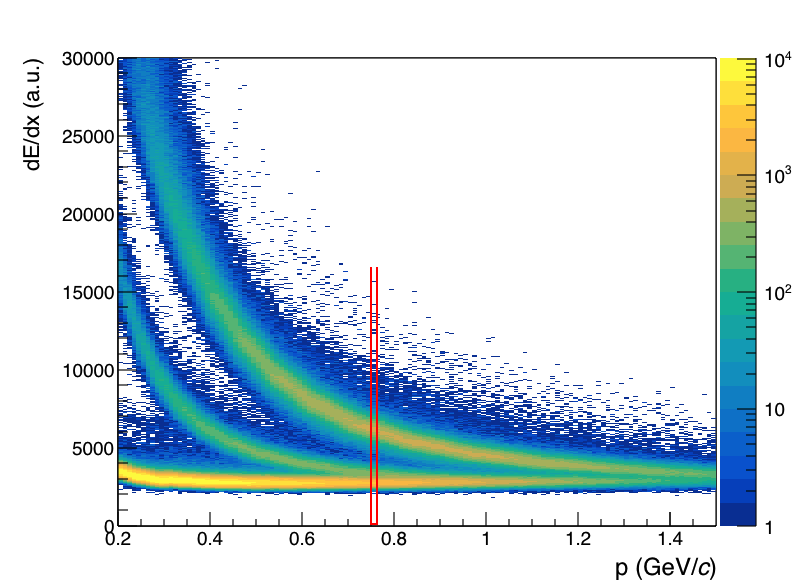} 
\put(41,62){ \color{darkgray} MPD Simulation }
\put(41,57){ \color{darkgray} TPC response}
\put(41,52){ \footnotesize \color{darkgray} Bi--Bi $\sNN=9.46$ GeV} 
\put(65,47){ \footnotesize \color{darkgray} (SMASH)}
\end{overpic}  
\begin{overpic}[width=0.48\textwidth, trim={0.1cm 0.0cm 0.3cm 0.3cm},clip]
{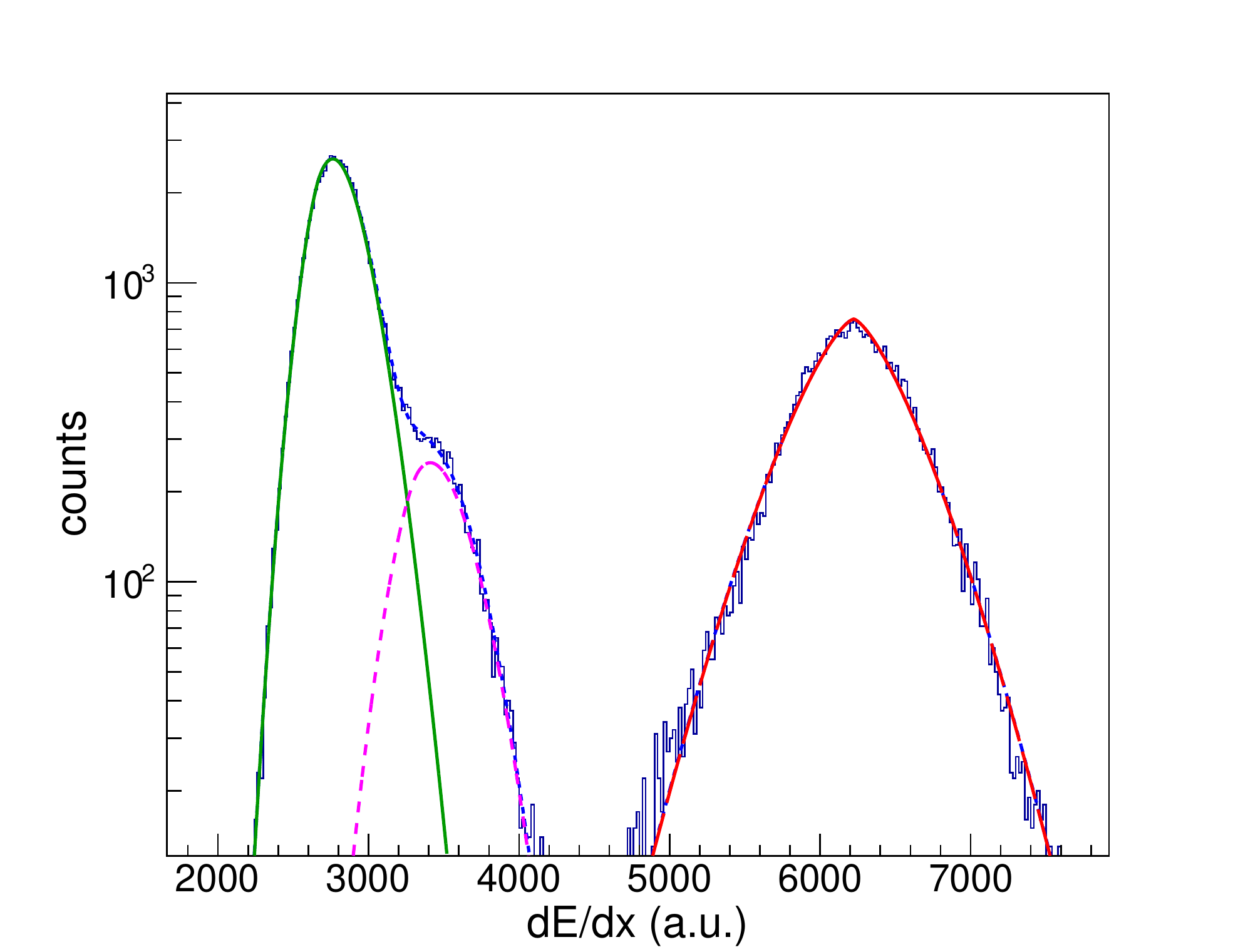} 
\put(31,62){\large \color{DarkGreen} $\pi^+$}
\put(39,38){\large \color{magenta} K$^+$}
\put(73,52){\large \color{red} p}
\put(43,66){ \footnotesize \color{darkgray}  $p\in(0.75-0.76)$ GeV/$c$} 
\end{overpic}  
\caption{
{\it Left:} distribution of the $dE/dx$ signal  in TPC versus track momentum.
{\it Right: } $dE/dx$ projection in the momentum bin 0.75-0.76 GeV/$c$ (denoted by a red box in the left panel),
with fits of underlying distributions of $\pi^+$, K$^+$, p yields  by Generalized Gaussians.
}
\label{QA_mom_bin}
\end{figure}

\section{Simulated dataset and $\mathbf{dE/dx}$ fits}

For the current study, we used a dataset of Bi--Bi collisions at $\sNN=9.46$ GeV simulated in SMASH event generator \cite{smash}, 
with subsequent 
simulation of the MPD detector response  in GEANT3,  digitization of the signals, 
and the full reconstruction of the events afterwards. 
Events with position of the vertex along the beam axis $z$ within $\pm$ 20 cm from the nominal interaction point 
were taken.
Centrality classes were selected by dividing the multiplicity distribution of the tracks observed in the TPC 
into quantiles, as it is typically done in other experiments.
Centrality class 0-20\% was used in this study, with 204$k$ events analyzed.

Tracks for the analysis were selected within pseudorapidity range $|\eta|<0.5$ 
and transverse momentum $\pt>0.15$ GeV/$c$.
Additionally, tracks were required to have minimum  30 clusters in TPC with $\chi^2$ per cluster less than 5,
and
the distance of closest approach to the primary vertex along $z$-axis less than 2 cm.
This selection allows one 
to reduce contamination by secondary particles from weak decays and detector material,
  keeping the efficiency at 80-90\% level. 

Left panel in Figure \ref{QA_mom_bin}
shows distribution of the $dE/dx$ signal  in TPC versus track momentum.
One may see typical trends
 for pions, kaons and protons (no electrons in SMASH), which start to overlap at high momenta.
In order to apply the Identity Method, 
one has to get $dE/dx$ projections in narrow momentum slices and fit them
by a sum of $\rho_i(dE/dx)$ functions. 
As an example,
a $dE/dx$ projection in the momentum bin 0.75-0.76 GeV/$c$
is shown in the right panel of Fig.\ref{QA_mom_bin}.
Fits for  $\pi^+$, K$^+$, p yields in this bin are done by the Generalized Gauss function,
 good quality of the overall fit is obtained. Similar fits are done for $dE/dx$ distributions
of the antiparticles ($\pi^-$, K$^-$, $\overline{p}$).
It is important to note 
that in the current work all fit parameters, except relative amplitudes between different species,
were obtained using the information about the true type of each particle,
which is available in simulations.
With real data,
if  a significant signal overlap between the species is present,
a simultaneous fitting for all particle species in a momentum slice is a challenging task. 
For that, one may use the so-called {\it clean samples} of particles with known PID,
which can be ``marked'' using products of V0 decays, taking TOF information, etc. \cite{Mesut_PhD}.

Fitted functions in all momentum slices and a ROOT tree containing $dE/dx$ and other event- and track-level
information
were  transferred then to the IM machinery.

\section{Reconstuction of  moments in single rapidity window}

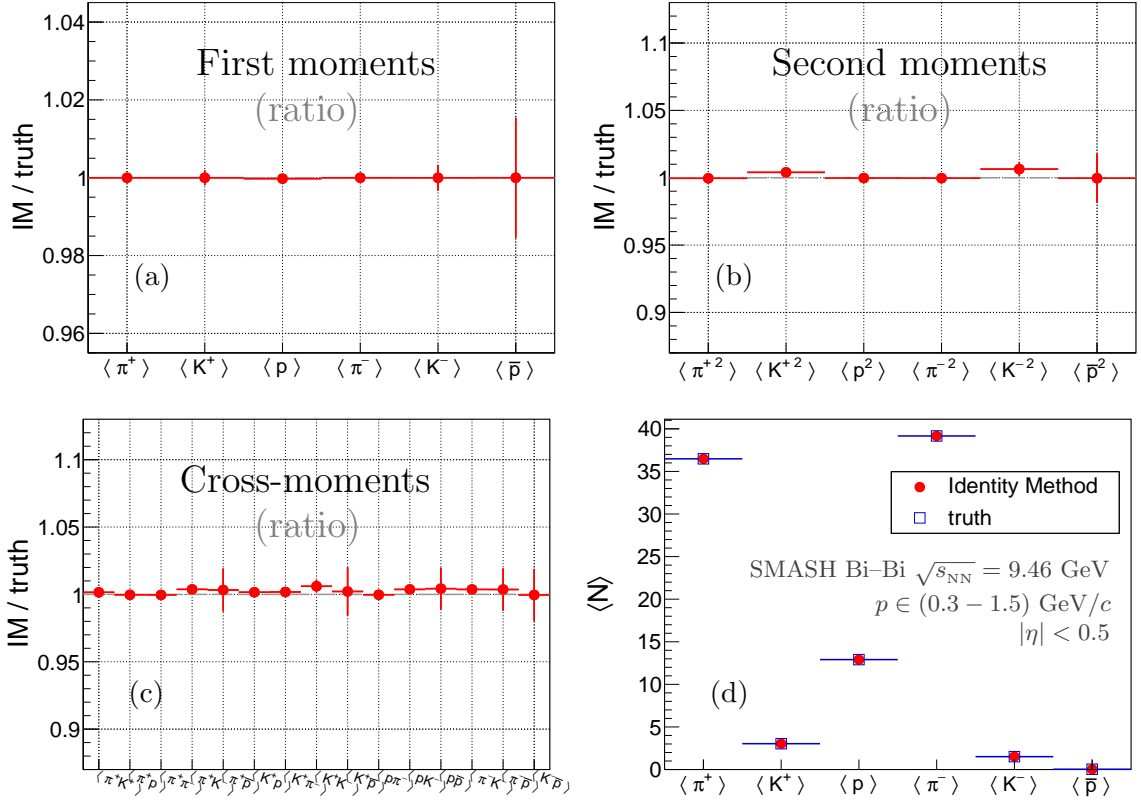
\begin{figure}[t]
\centering
\begin{overpic}[width=0.47\textwidth, trim={0.1cm 0.0cm 0.3cm 0.1cm},clip]
{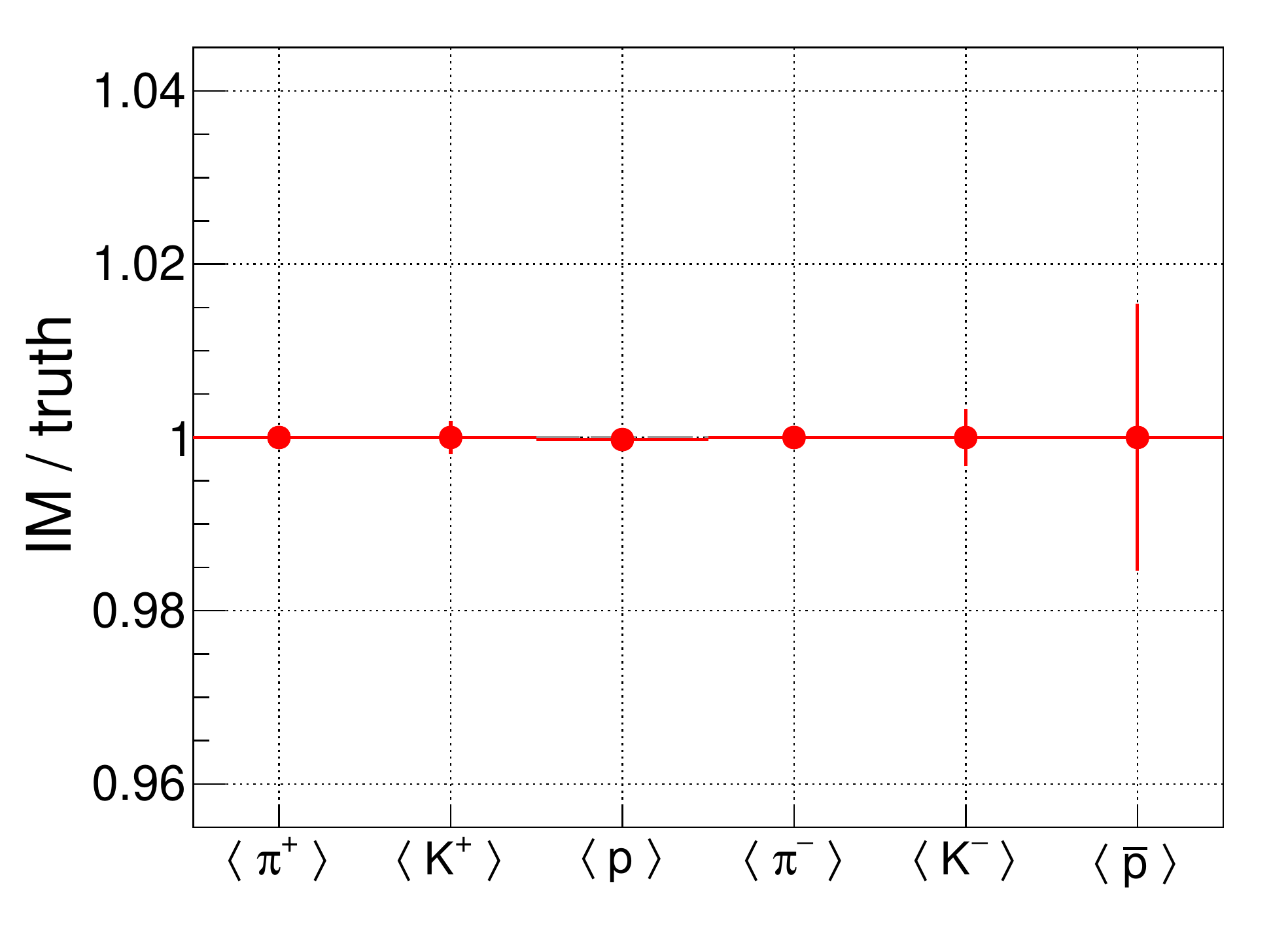} 
\put(21,20){ (a) }
\put(32,57){ \large First moments}
\put(42,49){ \color{gray} \large (ratio)}
\end{overpic}  
\begin{overpic}[width=0.47\textwidth, trim={0.1cm 0.0cm 0.3cm 0.1cm},clip]
{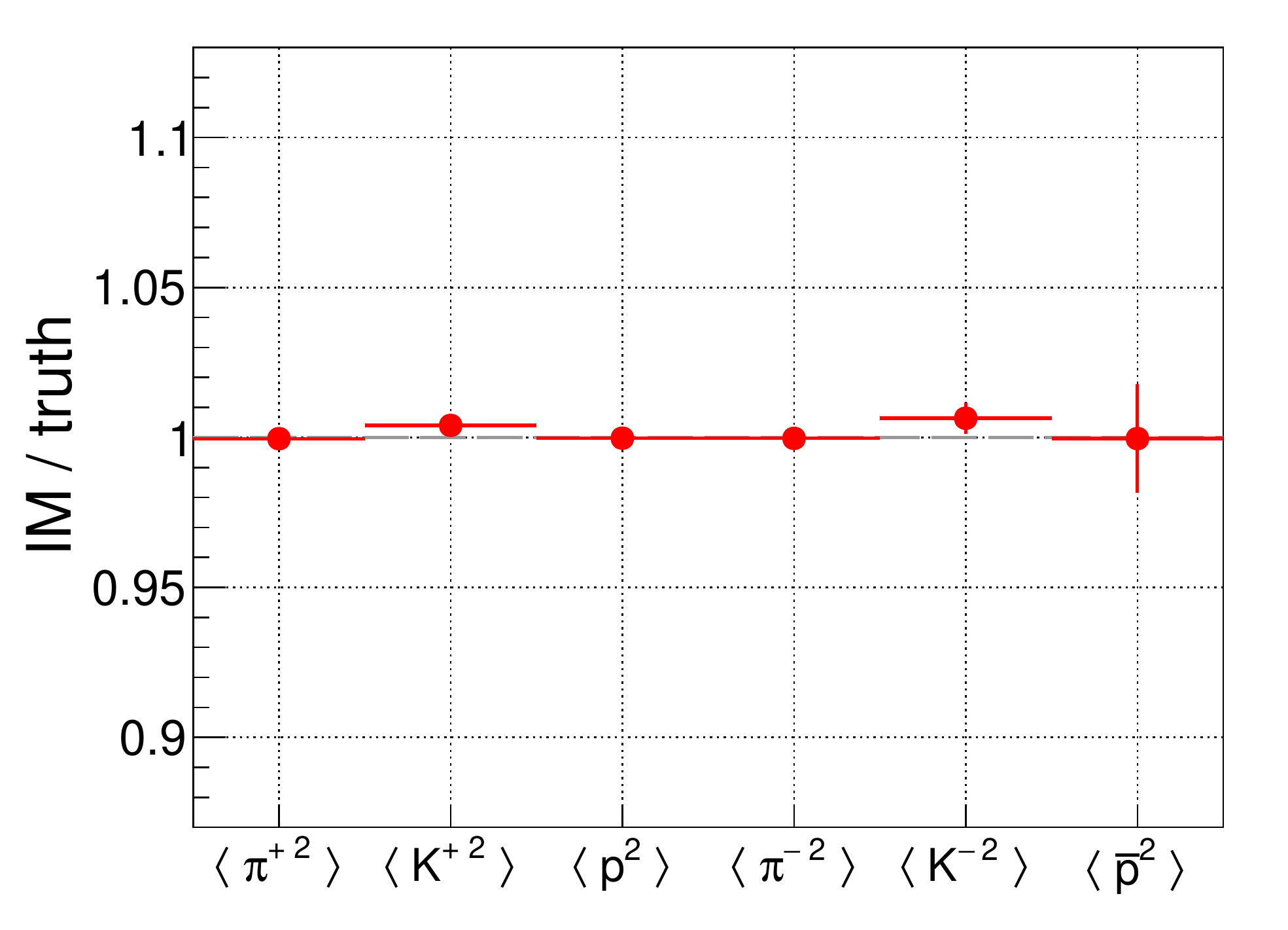} 
\put(21,20){ (b) }
\put(31,57){ \large Second moments}
\put(44,49){ \color{gray} \large (ratio)}
\end{overpic}  
\begin{overpic}[width=0.47\textwidth, trim={0.1cm 0.0cm 0.3cm 0.1cm},clip]
{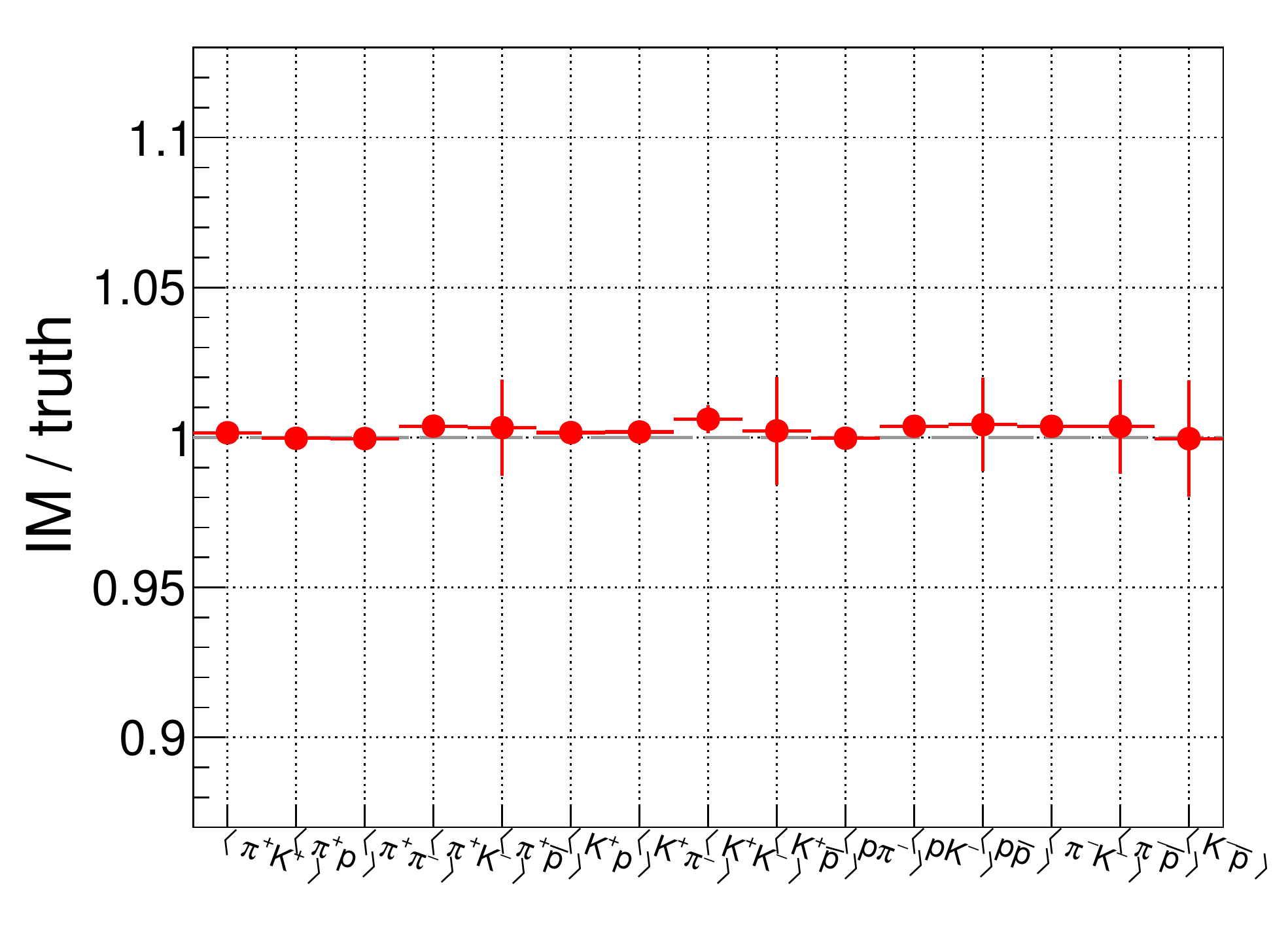} 
\put(21,20){ (c) }
\put(30,57){ \large Cross-moments}
\put(43,49){ \color{gray} \large (ratio)}
\end{overpic} 
\begin{overpic}[width=0.47\textwidth, trim={0.1cm 0.0cm 0.3cm 0.1cm},clip]
{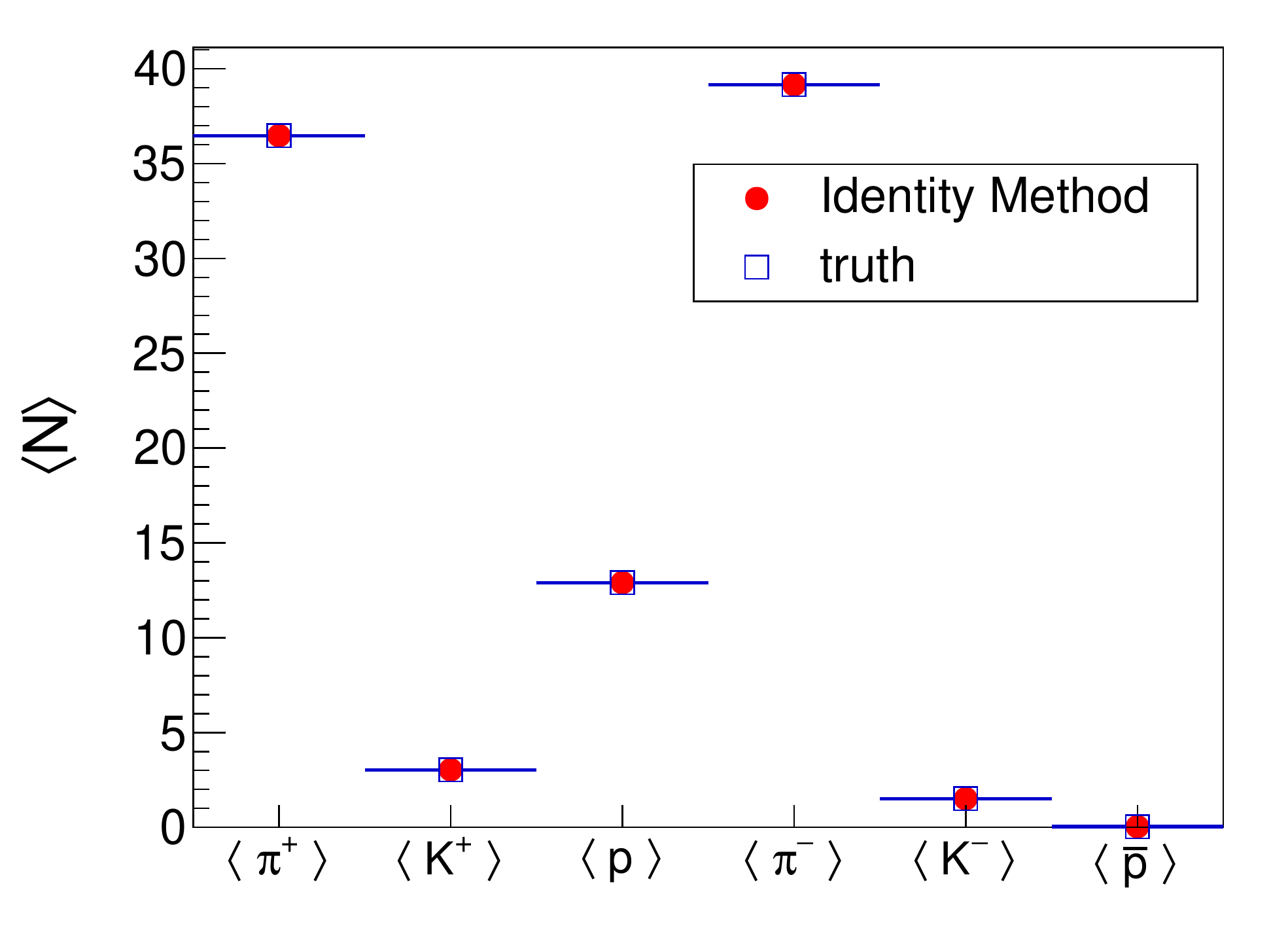} 
\put(21,20){ (d) }
\put(28,42){ \footnotesize \color{darkgray} SMASH Bi--Bi $\sNN=9.46$ GeV} 
\put(50,36){ \footnotesize \color{darkgray} $p\in(0.3-1.5)$ GeV/$c$} 
\put(75,31){ \footnotesize \color{darkgray} $|\eta|<0.5$} 
\end{overpic} \vspace{-2mm}
\caption{
Ratios of the  moments reconstructed with the Identity Method  to the true values,
for the first (a), second (b), and cross-moments (c).
Panel (d) shows the values of the first moments directly (i.e. before taking the ratio).
}
\label{moments_full_eta}
\end{figure}

At first, the analysis within a single rapidity window $|\eta|<0.5$ was performed.
The momentum range considered is  $(0.3,1.5)$ GeV/$c$.
The first ($\av{N_i}$, $i$ is a particle type), second ($\av{N_i^2}$), and cross-moments ($\av{N_i N_j}$) of identified particle yields were recovered with the IM.
Statistical uncertainties for the moments were calculated using the subsampling method (with 20 subsamples).
Performance of the Identity Method is shown in Figure \ref{moments_full_eta} (a--c)
in terms of the 
ratios of the IM-reconstructed moments   to the true values. 
The true moments were calculated using track PID that is explicitly known in simulations. 
It can be seen that good accuracy of the moment reconstruction is achieved (better than 1 \%).
The uncertainties are the largest for the moments that involve kaons and anti-protons, since their abundances at NICA energies are relatively low. This is visible in panel (d) of the  Fig.\ref{moments_full_eta}.


\section{ Forward-backward correlations of identified particle yields}

\begin{figure}[t]
\centering
\begin{overpic}[width=0.48\textwidth, trim={0.1cm 0.4cm 0.6cm 0.6cm},clip]
{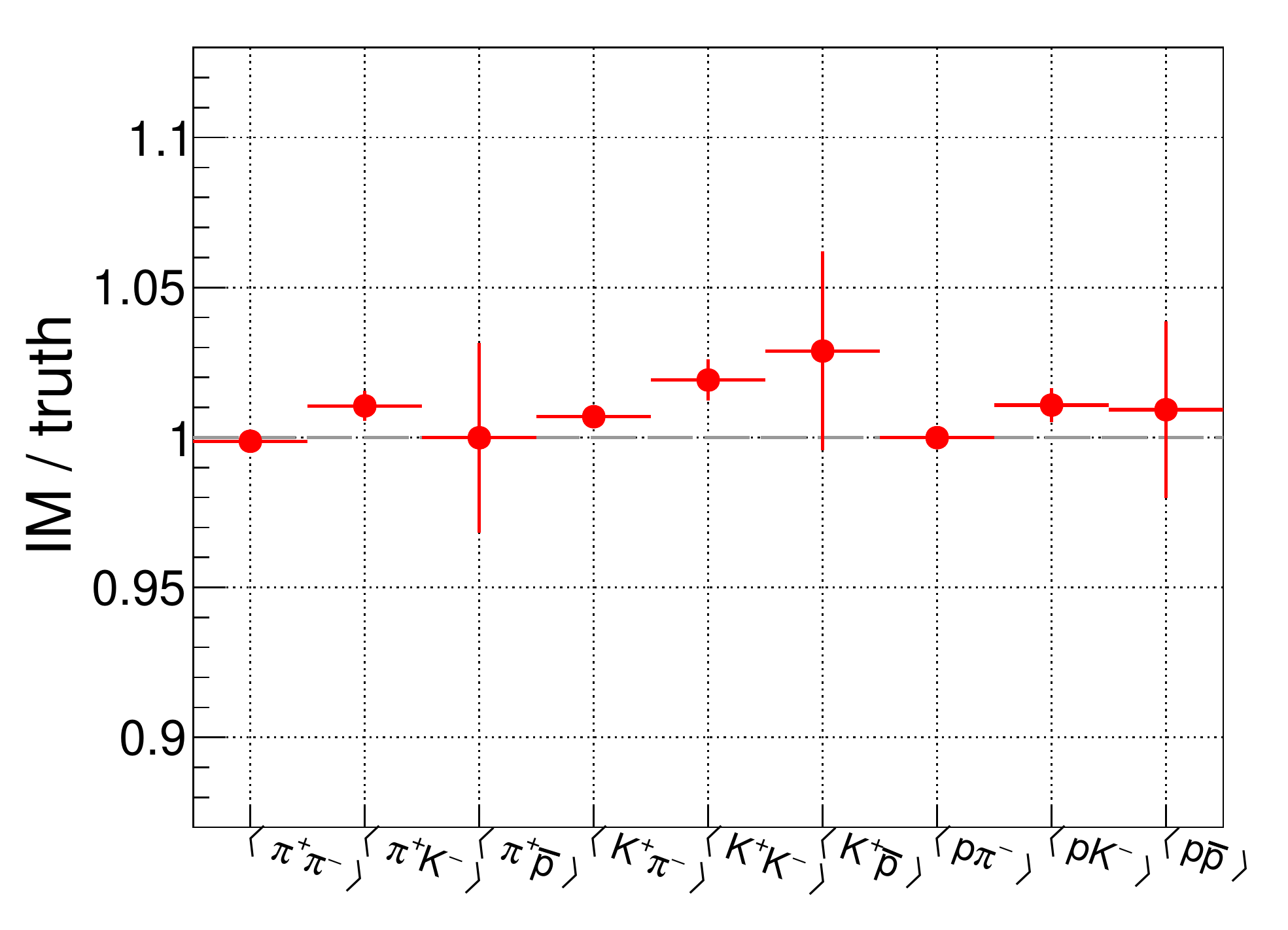} 
\put(27,56){ \large Cross F--B moments   }
\put(25,18){ \footnotesize \color{darkgray} $\eta_F\in(0.1, 0.5)$, $\eta_B\in(-0.5, -0.1)$ } 
\end{overpic}  
\caption{
Ratios of the cross-moments for 
 identified particle yields 
in forward and backward $\eta$-windows $(-0.5, -0.1)-(0.1, 0.5)$, recovered with the Identity Method,
to the true values.
}
\label{FB_results}
\end{figure}

With the Identity Method, one can also perform  more differential studies. Namely, we can select two regions in 
pseudorapidity and calculate the forward-backward  correlations between them,
for example, in terms of the Pearson correlation coefficient \cite{FB_Capella}:
\begin{equation}
\label{bcor}
b_{corr} = \frac{\av{N_{F,i} N_{B,j}}-\av{N_{F,i}}\av{N_{B,j}} } 
{ \sqrt{ \big(\av{N_{F,i}^2}-\av{N_{F,i}}^2  \big)  
\big(\av{N_{B,j}^2}-\av{N_{B,j}}^2}  \big) },
\end{equation}
where $F$ and $B$ denote the forward and the backward windows, respectively, and $i$, $j$ are particle species.
Figure \ref{FB_results} shows ratios of IM-reconstructed values to the true ones for 
the  cross-moments in a pair of FB windows,
where the forward window $\eta_F\in(0.1, 0.5)$ counts positive particles
and the backward window $\eta_B\in(-0.5, -0.1)$ -- negative particles.
A good closure of the ratios to unity is  observed 
(the biggest deviations $\sim$2-$3\%$ are, again, for kaons and $\overline{p}$ due to their small yields).
The correlation coefficient \eqref{bcor}  
extracted with the Identity Method
for the case of e.g.  $\pi^+ - \pi^-$ 
FB correlations is $b_{\rm corr}^{\rm IM}=0.357\pm 0.002$,
while the true value is $b_{\rm corr}^{\rm true} = 0.3598 \pm  0.0012$, the results coincide within the uncertainties.

\section{ Summary }
In this work,
performance of the Identity Method was tested on realistic A--A events with GEANT simulation and reconstruction in the MPD detector. The first results showed reasonable quality of the reconstructed moments.
Along with the conventional analysis in a single pseudorapidity window, 
it was suggested also to use the Identity Method for more differential studies,
in particular,  for forward-backward rapidity correlations.
It was shown that the Identity Method allows one to reliably
 reconstruct the true values of the FB correlation coefficient at MPD.

\section*{\bf Acknowledgements }
This work is supported by the RFBR research project No. 18-02-40097.


\end{document}